Title:
Data collaboration for causal inference from limited medical testing and medication data

Authors and affiliations:

Tomoru Nakayama
Graduate School of Science and Technology, University of Tsukuba, Tsukuba, Japan
s2420512@u.tsukuba.ac.jp

Yuji Kawamata (Corresponding author)
Center for Artificial Intelligence Research, University of Tsukuba, Tsukuba, Japan
yjkawamata@gmail.com

Akihiro Toyoda
Graduate School of Science and Technology, University of Tsukuba, Tsukuba, Japan
s2320513@u.tsukuba.ac.jp

Akira Imakura
Center for Artificial Intelligence Research, University of Tsukuba, Tsukuba, Japan
imakura@cs.tsukuba.ac.jp

Rina Kagawa
Artificial Intelligence Research Center, National Institute of Advanced Industrial Science and Technology, Tsukuba, Japan
sonata.skazka@gmail.com

Masaru Sanuki
Faculty of Medicine, Department of Clinical Medicine, University of Tsukuba, Tsukuba, Japan
sanuki@md.tsukuba.ac.jp

Ryoya Tsunoda
Faculty of Medicine, Department of Nephrology, University of Tsukuba, Tsukuba, Japan
tsunoda@md.tsukuba.ac.jp

Kunihiro Yamagata
Faculty of Medicine, Department of Nephrology, University of Tsukuba, Tsukuba, Japan
k-yamaga@md.tsukuba.ac.jp

Tetsuya Sakurai
Center for Artificial Intelligence Research, University of Tsukuba, Tsukuba, Japan
sakurai@cs.tsukuba.ac.jp

Yukihiko Okada
Center for Artificial Intelligence Research, University of Tsukuba, Tsukuba, Japan
okayu@sk.tsukuba.ac.jp

Abstract

Observational studies enable causal inferences when randomized controlled trials (RCTs) are not feasible. However, integrating sensitive medical data across multiple institutions introduces significant privacy challenges. The data collaboration quasi-experiment (DC-QE) framework addresses these concerns by sharing "intermediate representations"—dimensionality-reduced data derived from raw data—instead of the raw data. Although DC-QE can estimate treatment effects, its application to medical data remains unexplored. The aim of this study was to apply the DC-QE framework to medical data from a single institution to simulate distributed data environments under independent and identically distributed (IID) and non-IID conditions. We propose a method for generating intermediate representations within the DC-QE framework. Experimental results show that DC-QE consistently outperformed individual analyses across various accuracy metrics, closely approximating the performance of centralized analysis. The proposed method further improved performance, particularly under non-IID conditions. These outcomes highlight the potential of the DC-QE framework as a robust approach for privacy-preserving causal inferences in healthcare. Broader adoption of this framework and increased use of intermediate representations could grant researchers access to larger, more diverse datasets while safeguarding patient confidentiality. This approach may ultimately aid in identifying previously unrecognized causal relationships, support drug repurposing efforts, and enhance therapeutic interventions for rare diseases.

Introduction

Randomized controlled trials (RCTs) represent the "gold standard" for causal inference. However, conducting RCTs in epidemiological and medical contexts poses significant challenges due to ethical and financial constraints, particularly when timely interventions are required. These limitations render observational studies a valuable alternative for deriving causal inferences[1–3]. However, such studies rely on large datasets to ensure reliable outcomes, yet sharing raw data in sensitive domains such as medicine and finance remains challenging because of privacy concerns[4–7]. Hence, considerable research has focused on developing privacy-preserving approaches.

Federated learning (FL) is one such approach that has been widely adopted in machine-learning tasks[8]. FL achieves privacy preservation by sharing models instead of raw data. This methodology has garnered significant attention in the medical field[9], with numerous studies demonstrating its utility[10–12]. Specifically, Imakura and Sakurai introduced data collaboration (DC) analysis as a model non-sharing FL framework[13]. DC analysis facilitates data integration across horizontal and vertical dimensions without requiring continuous communication with external entities. An example of its application in healthcare includes collaboration between a hospital and a city to develop a model for diabetes prediction[14].

Kawamata et al. extended DC analysis by proposing the data collaboration quasi-experiment (DC-QE) framework, designed for causal inference[15]. DC-QE enables treatment effect estimation from distributed data while addressing covariate biases between treatment and control groups by combining DC analysis with propensity score matching (PSM)[16]. The increasing availability of diverse healthcare datasets (e.g., health checkup records, electronic medical records, and imaging data) coupled with the frequent use of PSM in actual cohort studies[17–20] underscores the potential of DC-QE in the medical field. However, the potential of DC-QE remains untested with medical data because it has been applied exclusively to artificial datasets and labor training programs.

In recent years, an increasing number of studies have explored causal inference methods based on FL. In many of these studies, the performance of their proposed methods was evaluated using medical (or healthcare) data. For example, they examined the effects of early intervention on children's cognitive development[21], medication for patients with acute respiratory distress[22], and surgery for those with acute myocardial infarction[23]. In contrast, as mentioned previously, the performance of DC-QE has not yet been assessed using medical data. Investigating DC-QE in a medical context could provide valuable insights into its practical applications.

In this study, the effectiveness of DC-QE was evaluated using medical data. Specifically, the evaluation focused on assessing the DC-QE framework for estimating the effect of uric acid-lowering treatments on reducing uric acid levels. Kawamata et al.[15] conducted experiments under independent and identically distributed (IID) data distribution scenarios. Therefore, we initiated experiments under IID conditions. Subsequently, we conducted experiments under non-IID conditions to extend the evaluation to more generalized settings, where local data distributions were non-IID, because medical data usually have different distributions by region or hospitals [24], and previous studies[22,23] have evaluated their methods only for non-IID situations. Additionally, we propose a method for generating intermediate representations, a dimensionality reduction technique aimed at enhancing DC-QE performance. The main contributions of this study are as follows:

- A practical dimensionality reduction method for DC-QE is proposed.
- The effectiveness of DC-QE on medical data partitioned into IID and non-IID settings is demonstrated for the first time.
    - Conducting full collaboration among all institutions using DC-QE under IID settings resulted in performance surpassing individual analyses across most evaluation metrics.
    - Selecting appropriate intermediate representations enabled DC-QE to outperform individual analyses across multiple evaluation metrics under non-IID settings.

## Data and methodology

The handling of medical data was approved by the Ethical Review Committee for Clinical Research of the University of Tsukuba Hospital (permission number: R04-147). All procedures involving human participants were conducted in accordance with the ethical standards of the 1964 Helsinki Declaration and its later amendments. Informed consent was not obtained for this study because it did not involve intervention or the collection of new samples. Instead, opt-outs were obtained by posting a written explanation of the study in the hospital and on the website, including the possibility of refusing to allow medical information to be used in the study, according to the Ethical Guidelines for Medical and Biological Research Involving Human Subjects established by the Ministry of Health, Labor and Welfare of Japan (https://www.mhlw.go.jp/content/001077424.pdf, in Japanese). The need to obtain informed consent was waived by the Ethics Committee of University of Tsukuba Hospital (IRB Approval Number R04-147).

## Study design and data settings

The study used examination and prescription history data from Tsukuba University Hospital, spanning 2014 to 2020. The examination history dataset included blood tests and urinalysis data, and the prescription history dataset documented medications prescribed to patients during this period.

The experiment focused on verifying the effectiveness of uric-lowering agents (ULAs), specifically allopurinol and febuxostat (hereafter, Target ULAs), in reducing serum uric acid (SUA) levels. Covariates were selected on the basis of a study by Chou et al.[25]. Given that previous studies have demonstrated a significant reduction in SUA levels within two weeks of Target ULA administration[26-28], in this study we examined reductions in SUA levels between two and four weeks following prescription. Although Chou et al. included allopurinol, febuxostat, and benzbromarone as treatment targets, benzbromarone was excluded in the present study because of insufficient sample size in the dataset.

The exclusion criteria for study subjects were as follows:
- **(i)** Age < 20 or > 100 years.
- **(ii)** History of Target ULA prescriptions prior to July 2014.
- **(iii)** Use of multiple ULAs during the observation period.
- **(iv)** Absence of baseline or follow-up SUA data.
- **(v)** Treatment-to-follow-up SUA test interval shorter than two weeks or longer than one month.
- **(vi)** Missing values for over half of the covariates, excluding prescription history.

Features with missing values in more than half of the instances were removed after applying criterion (v). Criterion (iii) functioned as a preprocessing step to exclude the influence of medications other than the target ULAs on SUA levels. The ULAs considered for criterion (iii) included allopurinol, febuxostat, topiroxostat, benzbromarone, probenecid, dotinurad, and rasburicase. After preprocessing, missing values were imputed using the median values. Fig. 1 illustrates the preprocessing workflow, and Table 1 provides a summary of the statistical properties of the dataset after preprocessing.

The original dataset lacked values for the estimated glomerular filtration rate (eGFR), which was calculated using the following equation: $eGFR = 194 \times SCr^{-1.094} \times Age^{-0.287} \times sex$, where $Age$ denotes the patient's age, $SCr$ refers to the serum creatinine level, and $sex$ was assigned a value of 1 for males and 0.739 for females[29]. Baseline eGFR values were computed using the most recent laboratory results available prior to treatment for the treatment group and prior to uric acid measurement for the control group. Chronic kidney disease (CKD) stages were categorized on the basis of eGFR values into the following intervals: >90, 60–90, 30–60, 15–30, and <15 mL/min/1.73 m$^2$.

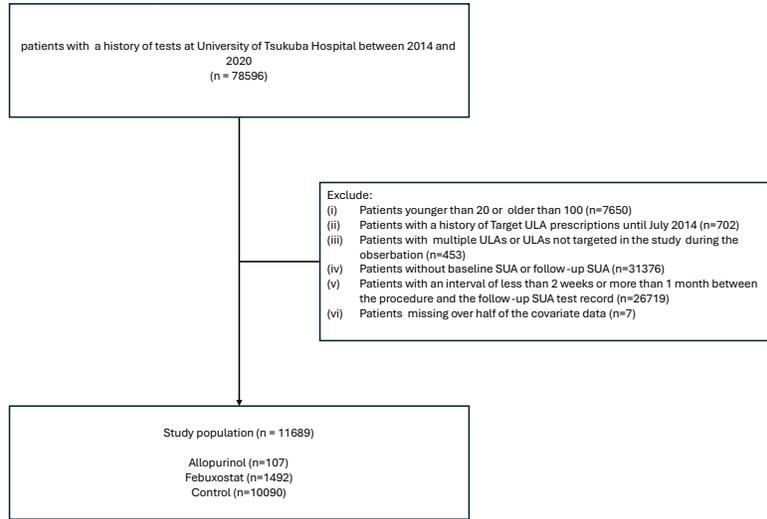

**Fig. 1.** Preprocessing flowchart, where *n* indicates the number of samples.

**Table 1**. Basic statistics after preprocessing.

| Sample size | unit | Control n = 10090 | | Allopurinol n = 107 | | Febuxostat n = 1492 | |
|---|---|---|---|---|---|---|---|
| | | N | % | N | % | N | % |
| Diuretic | - | 33 | 0.33 | 7 | 6.54 | 86 | 5.76 |
| Beta-adrenergic antagonist | - | 167 | 1.66 | 15 | 14.02 | 109 | 7.31 |
| Renin-angiotensin system inhibitor | - | 182 | 1.80 | 19 | 17.76 | 191 | 12.80 |
| Antipurine | - | 9 | 0.09 | 2 | 1.87 | 12 | 0.80 |
| Alkalizer | - | 174 | 1.72 | 5 | 4.67 | 153 | 10.25 |
| Nonsteroidal anti-inflammatory drug | - | 259 | 2.57 | 9 | 8.41 | 105 | 7.04 |
| Aspirin | - | 10 | 0.10 | 2 | 1.87 | 9 | 0.60 |
| Corticosteroid | - | 422 | 4.18 | 13 | 12.15 | 257 | 17.23 |
| Colchicine | - | 2 | 0.02 | 0 | 0.00 | 5 | 0.34 |
| CKD stage 1 | - | 4056 | 40.20 | 8 | 7.48 | 189 | 12.67 |
| 2 | - | 2483 | 24.61 | 19 | 17.76 | 233 | 15.62 |
| 3 | - | 2777 | 27.52 | 55 | 51.40 | 558 | 37.40 |
| 4 | - | 341 | 3.38 | 6 | 5.61 | 234 | 15.68 |
| 5 | - | 431 | 4.27 | 18 | 16.82 | 278 | 18.63 |
| | | mean | std | mean | std | mean | std |
| High-density lipoprotein cholesterol | mg/dl | 52.55 | 17.20 | 49.03 | 16.10 | 43.98 | 15.77 |
| Hemoglobin A1c | % | 6.16 | 1.38 | 6.05 | 1.13 | 6.09 | 1.10 |
| eGFR | g/dl | 86.61 | 49.99 | 46.51 | 28.62 | 48.80 | 37.34 |
| Albumin | mg/dl | 3.83 | 0.69 | 3.79 | 0.74 | 3.59 | 0.76 |
| Creatinine | mg/dl | 1.01 | 1.41 | 1.91 | 2.26 | 2.06 | 2.38 |
| Triglycerides | mg/dl | 134.82 | 116.51 | 163.46 | 113.58 | 157.02 | 127.13 |
| Serum uric acid | mg/dl | 5.09 | 1.77 | 6.01 | 2.32 | 7.00 | 2.70 |
| Age | - | 58.09 | 17.47 | 65.57 | 13.37 | 62.12 | 15.03 |
| Sex | - | 0.52 | 0.50 | 0.16 | 0.37 | 0.30 | 0.46 |
| Hemoglobin | g/dl | 12.59 | 2.26 | 12.43 | 2.59 | 11.38 | 2.65 |
| Urine pH | - | 6.50 | 0.79 | 6.30 | 0.90 | 6.25 | 0.75 |

| | | | | | | |
|---|---|---|---|---|---|---|
| Urine occult blood | - | 0.76 | 1.30 | 0.65 | 1.13 | 0.96 | 1.33 |

## Data collaboration quasi-experiment

In this study, the DC-QE framework played two distinct roles: data users and analysts. Users provided data, and analysts were responsible for processing and analyzing the shared information. The algorithm comprised three stages: Each user retained a private dataset $X_k \in \mathbb{R}^{n_k \times m}$, a treatment variable $Z_k$, and an outcome variable $Y_k$, where $k$ denotes the identifier of a user, $n_k$ represents the number of samples of user $k$, and $m$ indicates the number of covariates.

Instead of sharing raw data $X_k$, users transform their datasets into intermediate representations suitable for sharing. The intermediate representations provided by users were integrated by the analyst into a unified collaborative representation that enabled the construction of a propensity score model. The analyst estimated the propensity scores and applied the matching results to compute the average treatment effect on the treated $\tau_{ATT}^{DC-QE(PSM)}$ using the collaborative representation.

These steps outline the DC-QE process, facilitating collaborative data analysis while maintaining privacy. In this study, we specifically examined horizontal data partitioning scenarios. Further algorithmic details can be found in the work of Kawamata et al.[15].

### *First step*

In the first step, each user generated an intermediate representation of that user's dataset and shared it with the analyst. A common dataset, called anchor data $X^{anc} \in \mathbb{R}^{r \times m}$, was required for constructing intermediate representations, where $r$ denotes the number of samples. Anchor data can be derived from publicly available or basic statistical datasets using techniques such as random sampling, low-rank approximation, or SMOTE[30,31]. These anchor data were essential for creating collaborative representations.

Each user produced an intermediate representation $\tilde{X}_k$ using that user's dataset $X_k$ and applied a transformation function $f_k$ to $\tilde{X}_k$. The intermediate representations are expressed as

$$\tilde{X}_k = f_k(X_k) \in \mathbb{R}^{n_k \times \tilde{m}_k} \quad (1)$$
$$\tilde{X}_k^{anc} = f_k(X^{anc}) \in \mathbb{R}^{r \times \tilde{m}_k} \quad (2)$$

where $f_k$ represents a row-wise dimensionality reduction function, which can include methods such as principal component analysis (PCA)[32], linear discrimination analysis[33], and locality-preserving projections (LPP)[34]. The number of dimensions after reduction, $\tilde{m}_k$, was determined independently by each user. Once the intermediate representations were constructed, the representations $\tilde{X}_k$, $\tilde{X}_k^{anc}$, along with the treatment $Z_k$, and outcome $Y_k$, were shared with the analyst.

### *Second step*

In the subsequent step, the analyst constructed collaborative representations from the intermediate representations provided by the users. A transformation function must be applied to standardize these representations for integrated analysis because the dimensions of the shared intermediate representations may vary across users. For all $k$, the transformation function satisfied

$$g_k(\tilde{X}_k^{anc}) \approx g_{k'}(\tilde{X}_{k'}^{anc}) \in \mathbb{R}^{r \times \breve{m}} \ (k \neq k') \quad (3)$$

where $\breve{m}$ denotes the dimension of the collaborative representation. Assuming $g_k$ is a linear transformation, the transformation function $g_k$ for user $k$ can be defined as

$$\breve{X}_k = g_k(\tilde{X}_k) = \tilde{X}_k G_k \in \mathbb{R}^{n_k \times \breve{m}} \quad (4)$$

where $G_k$ represents the linear transformation matrix. The collaborative representation using the computed $G_k$ is expressed as

$$\breve{X} = \begin{bmatrix} \breve{X}_1 \\ \breve{X}_2 \\ \vdots \\ \breve{X}_c \end{bmatrix} = \begin{bmatrix} \tilde{X}_1 G_1 \\ \tilde{X}_2 G_2 \\ \vdots \\ \tilde{X}_c G_c \end{bmatrix} \in \mathbb{R}^{n \times \breve{m}} \quad (5)$$

*Third step*

In the third step, a model was developed to estimate the propensity score based on the collaborative representation, and the treatment effect was calculated using PSM. The propensity score $\alpha$ is defined as

$$\alpha_i = Pr(z_i = 1|\breve{x}_i) \quad (6)$$

where $\breve{x}_i$ represents the collaborative representation of covariates for subject $i$. Logistic regression is commonly employed to estimate propensity scores. The matching pair $pair^{DC-QE}(i)$ for subject $i$ was determined using these estimated propensity scores in the DC-QE framework. The treatment effect is estimated by

$$\hat{\tau}_{ATT}^{DC-QE(PSM)} = \frac{1}{n}\sum_{i\in\mathbb{N}_T}(y_i - y_{pair^{DC-QE}(i)}) \quad (7)$$

*Privacy preservation in DC-QE*

Here we provide a privacy analysis of DC-QE. The DC frameworks have been described as having a double privacy layer for private data $X_k$[35]:

- First layer: No one can possess private data $X_k$ because $F_k$ is not shared. In particular, $F_k$ can be discarded for each user.
- Second layer: Even if $F_k$ is stolen, the private data $X_k$ is still protected for $\epsilon$-DR privacy [36] because $F_k$ is a dimensionality reduction function.

The first layer is based on the fact that $F_k$ remains private and cannot be deduced by others, because the input and output of $F_k$ are not publicly accessible. Consequently, no one can reconstruct $X_k$ only from the shared intermediate representation $\tilde{X}_k = X_k F_k$. The second layer is based on the fact that $F_k$ is a dimensionality reduction function. Therefore, the original data $X_k$ cannot be obtained from $\tilde{X}_k = X_k F_k$. Imakura et al. [37] experimentally evaluated the reconstruction of $X_k$ form $\tilde{X}_k$. Note that statistics could be leaked from the anchor data, which represents a potential negative impact of the proposed method.

*Dimensionality reduction method for DC-QE*

The matching performance in DC-QE was influenced by the dimensionality reduction method. Therefore, we propose a bootstrap-based dimensionality reduction method, inspired by the work of Imakura et al.[37], to enhance this performance. This approach estimates multiple parameters using the bootstrap method and constructs a subspace from the resulting values. Let $\boldsymbol{\beta}_k$ denote the model parameters that estimate the propensity scores computed from the subspace $\mathcal{S}_k = \mathcal{R}(F_k G_k) \subset \mathcal{R}(F_k)$. If the range $\mathcal{R}(F_k)$ includes the true estimator derived from centralized analysis (CA), it can also be approximated through DC-QE. The bootstrap-based method provides a robust alternative because $\beta_k$, estimated from the dataset of each user, approximates the centralized true estimator. Given a sampling ratio $0 < p < 1$, user $k$ randomly extracted $pn_k$ samples from that user's total dataset of size $n_k$. The model parameters were estimated through this sampling process, which was repeated $\tilde{m}_{BS}$ times to generate $\boldsymbol{\beta}_1, \boldsymbol{\beta}_2, \ldots, \boldsymbol{\beta}_{\tilde{m}_{BS}}$. The constructed parameters formed the matrix: $F_k^{BS} = [\boldsymbol{\beta}_1, \boldsymbol{\beta}_2, \ldots, \boldsymbol{\beta}_{\tilde{m}_{BS}}] \in \mathbb{R}^{m_k \times \tilde{m}_{BS}}$. The bootstrap function (BS), which was combined with another dimensionality reduction function $F^{DR} \in \mathbb{R}^{m_k \times \tilde{m}_{DR}}$, was generated using alternative methods, where $\tilde{m} = \tilde{m}_{BS} + \tilde{m}_{DR}$. The combined bootstrap-based dimensionality reduction function $F_k$ is defined as $F_k = [F_k^{DR}, F_k^{BS}]E_k$, where $E_k \in \mathbb{R}^{\tilde{m} \times \tilde{m}}$ is a randomly generated orthogonal matrix. The process of bootstrap-based dimensionality reduction for DC-QE is summarized in Algorithm 1.

| Algorithm 1 A bootstrap-based dimensionality reduction for DC-QE |
|---|
| Input: $X \in \mathbb{R}^{n \times m}$ and $Y \in \mathbb{R}^n$, and parameters $\tilde{m} = \tilde{m}_{bs} + \tilde{m}_{dr}$ and $p$ |
| Output: $F \in \mathbb{R}^{n \times \tilde{m}}$ |
| 1:  for $s = 1, 2, \ldots, \tilde{m}_{bs}$ do |
| 2:    Set $X', Y'$ by a random sampling of size $pn$ |
| 3:    Apply the logistic regression model to $X'$ and $Y'$ to obtain $\boldsymbol{\beta}_s$ |

4:    end for
5:    Set $F^{BS} = [\boldsymbol{\beta}_1, \boldsymbol{\beta}_2, \ldots, \boldsymbol{\beta}_{\widetilde{m}_{bs}}]$
6:    Apply dimensionality reduction to $X', Y'$ to obtain $F^{DR} = [\boldsymbol{\beta}_1, \boldsymbol{\beta}_2, \ldots, \boldsymbol{\beta}_{\widetilde{m}_{dr}}]$
7:    Set $F = [F^{BS}, F^{DR}]E$

## Experiments

### Common settings and evaluation scheme

The performance of DC-QE was compared against CA and individual analysis (IA). CA represents the ideal scenario where all raw data of users can be shared, including all covariates $X = [X_1, X_2, \ldots, X_c]$, treatments $Z = [Z_1, Z_2, \ldots, Z_c]$, and outcomes $Y = [Y_1, Y_2, \ldots, Y_c]$, enabling a comprehensive estimation of treatment effects. However, IA estimated treatment effects using only the dataset owned by a single user because of privacy constraints. Specifically, IA used only the covariates $X_k$, treatment $Z_k$, and outcome $Y_k$ associated with user $k$.

The experiments were designed to verify whether DC-QE outperformed IAs and provided results comparable to those achieved through CA. We created pseudo-user groups by partitioning the data in two ways to simulate the data distribution: IID and non-IID partitioning. Details of these partitioning methods are provided in subsequent sections. Each experiment employed the bootstrap method to randomly sample all data samples and estimate the treatment effect from the extracted datasets. The number of bootstrap repetitions $B$ was fixed at 1,000. We note that the bootstrap method in these experiments involves sampling from the population under study and should not be conflated with the bootstrap-based dimensionality reduction technique used by individual users to construct intermediate representations. Logistic regression was used to estimate propensity scores, and caliper matching was applied for the matching process. Note that PSM is a highly popular causal estimator used to estimate a treatment effect from observational data [38-40]. Thus, as a first step to show the applicability of DC-QE, we used PSM in our study.

The parameters used in DC-QE were set as follows: Anchor data were generated by random sampling from a uniform distribution bounded by the minimum and maximum values of each covariate. The sample size of the anchor data was defined as $r = \Sigma_{k=1}^{C} n_k$. The dimension of the intermediate and collaborative representations was configured as $\breve{m} = \widetilde{m} = m - 1$.

Experiments were conducted with the following three combinations for the dimensions generated using the bootstrap-based method $\widetilde{m}_{BS}$ and those created through dimensionality reduction $\widetilde{m}_{DR}$: $(\widetilde{m}_{DR}, \widetilde{m}_{BS}) = (\widetilde{m}, 0), \left(\frac{\widetilde{m}}{2}, \frac{\widetilde{m}}{2}\right), (0, \widetilde{m})$. Logistic regression was employed for bootstrap-based construction. PCA, LPP, and autoencoder (AE) [41] were used for dimensionality reduction. These configurations evaluated the performance of DC-QE under bootstrap-based dimensionality reduction.

The evaluation metrics included inconsistency, maximum absolute standardized mean difference (MASMD) and gap, with additional metrics for non-IID distributions, such as maximum absolute Jeffreys divergence (MJD), to assess whether the distribution after matching approximated the true distribution. Smaller values for these metrics indicate better performance. Each evaluation metric is explained below.

*Accuracy of estimated treatment effect*
Significant deviation between the estimated treatment effect and the true value indicates the ineffectiveness of the algorithm. Therefore, we used the gap method to measure the deviation between estimated treatment effects and those obtained through CA. The gap is calculated by

$$Gap(\hat{\boldsymbol{\tau}}, \hat{\boldsymbol{\tau}}^{CA}) = \sqrt{\frac{1}{B}\Sigma_{b=1}^{B}(\hat{t}_b - \hat{t}_b^{CA})^2} \quad (8)$$

where $B$ represents the number of bootstrap iterations, while $\hat{\boldsymbol{\tau}} = [\hat{\tau}_1, \hat{\tau}_2, \ldots, \hat{\tau}_B]$ and $\hat{\boldsymbol{\tau}}^{CA} = [\hat{\tau}_1^{CA}, \hat{\tau}_2^{CA}, \ldots, \hat{\tau}_B^{CA}]$ denote the estimated treatment effects obtained during each bootstrap iteration.

*Inconsistency of estimated propensity score*

Propensity scores play a critical role in PSM. Significant deviations between estimated propensity scores and their true values result in suboptimal matching performance. Inconsistency was evaluated by comparing the propensity scores from CA ($\hat{\boldsymbol{e}}^{CA} = [\hat{e}_1^{CA}, \hat{e}_2^{CA}, \ldots, \hat{e}_n^{CA}]$) with scores obtained in the experiments ($\hat{\boldsymbol{e}} = [\hat{e}_1, \hat{e}_2, \ldots, \hat{e}_n]$) as follows:

$$Inconsistency(\hat{\boldsymbol{e}}, \hat{\boldsymbol{e}}^{CA}) = \sqrt{\frac{1}{n}\sum_{i=1}^{n}(\hat{e}_i - \hat{e}_i^{CA})^2} \quad (9)$$

*Covariate balance*

Matching aims to balance covariates between treatment and control groups, making covariate balance an essential evaluation metric. The standardized mean difference (SMD) is commonly used for this purpose[42] and is expressed by

$$d_{cont}^{j} = \frac{\bar{x}_T^j - \bar{x}_C^j}{\sqrt{(s_T^j + s_C^j)/2}}, \quad d_{bin}^{j} = \frac{\hat{p}_T^j - \hat{p}_C^j}{\sqrt{(\hat{p}_T^j(1-\hat{p}_T^j) + \hat{p}_C^j(1-\hat{p}_C^j))/2}} \quad (10)$$

where $d_{cont}^{j}$ and $d_{bin}^{j}$ represent the SMDs for continuous and binary covariates $j$, respectively. $\bar{x}_T^j$ and $\bar{x}_C^j$ denote the mean covariate $j$ in the treatment and control groups, respectively. Similarly, $\bar{s}_T^j$ and $\bar{s}_C^j$ represent the corresponding standard deviations, and $\hat{p}_T^j$ and $\hat{p}_C^j$ denote the proportions of covariate $j$ equal to 1 in the treatment and control groups, respectively.

We evaluated covariate balance using the MASMD, which is defined as the maximum absolute value of the SMDs across all covariates, expressed by

$$MASMD(\boldsymbol{d}) = \max_j |d^j| \quad (11)$$

where $\boldsymbol{d} = [d_{cont}^1, \ldots, d_{cont}^{m_{cont}}, d_{bin}^1, \ldots, d_{bin}^{m_{bin}}]$.

*Distribution similarity between two groups*

The distribution obtained after matching may not accurately reflect the true distribution in non-IID settings, where the covariate distributions of data differ across users. Although MASMD effectively evaluates the covariate balance between treatment and control groups, it does not measure the fidelity of the distribution post-matching. Therefore, assessing the alignment between the covariate distribution and the true distribution after matching is crucial. Jeffreys divergence[43], a symmetrized version of the Kullback-Leibler divergence, was employed to evaluate the similarity between the two distributions:

$$D_J(P^j \| P_{CA}^j) = KL(P^j \| P_{CA}^j) + KL(P_{CA}^j \| P^j) \quad (12)$$

$$KL(P^j \| P_{CA}^j) = \sum_{b=1}^{bins} p^j(b) \log \frac{p^j(b)}{p_{CA}^j(b)} \quad (13)$$

where $P^j$ and $P_{CA}^j$ represent the probability distributions of covariate $j$. The variable "bins" denotes the number of intervals in the probability distribution, set to 20 for this experiment. We evaluated the Jeffreys divergence using its maximum value:

$$MJD = \max_j D_J(P^j \| P_{CA}^j) \quad (14)$$

Moreover, we assumed that the covariate distribution obtained under the CA setting was the true distribution for experimental purposes because the true distribution was unknown.

Data partitioning

The data-partitioning methods are explained for both IID and non-IID settings. Data in the IID configuration were randomly divided into 50 parts, simulating a scenario with 50 users (e.g., small hospitals). The performance of DC-QE was evaluated with varying numbers of collaborating users (2, 5, 10, 30, and 50). Experimental results for each case were compared with those from IA and CA.

Data partitioning in a non-IID scenario was designed to simulate three hospitals, employing the following methods:
- **Quantity**: Data were partitioned such that each user held 20%, 30%, or 50% of the total data, allocated randomly.
- **Label ratio**: Data were divided such that each user held an equal number of samples, with the proportion of treatment group samples for each user constituting 0.5%, 1%, or 2% of the total data samples.
- **Cluster**: Data were grouped into three clusters using uniform manifold approximation and
- projection and k-means clustering based on data covariates.

Performance evaluations for IA, CA, and DC-QE were compared in all non-IID scenarios. Notably, cluster-based partitioning is expected to result in more substantial covariate distribution differences across institutions compared with the quantity and label ratio methods because it is driven by covariate characteristics.

## Results

### Matching result

A specific example is presented in Table 2 to demonstrate the reduction in covariate bias achieved through DC-QE matching. It shows the SMDs pre- and post-matching. This example involves an experiment where allopurinol was used as the treatment. The DC-QE results correspond to the scenario in which PCA was employed to generate the intermediate representation, and collaboration occurred across all users. This is a single example, and comparisons between IA and CA are provided in subsequent sections. Table 2 indicates that matching with DC-QE reduced bias to a certain extent.

**Table 2.** Matching example of DC-QE. Data are presented as mean (std).

| Sample size | unit | Before matching | | | After matching (DC-QE) | | |
|---|---|---|---|---|---|---|---|
| | | $n$ = 10197 | | | $n$ = 234 | | |
| | | Control | Treatment | SMD | Control | Treatment | SMD |
| Diuretic | - | 0.003 (0.057) | 0.065 (0.248) | 0.346 | 0.043 (0.203) | 0.051 (0.222) | 0.04 |
| Beta-adrenergic antagonist | - | 0.017 (0.128) | 0.14 (0.349) | 0.473 | 0.068 (0.253) | 0.094 (0.293) | 0.094 |
| Renin-angiotensin system inhibitor | - | 0.018 (0.133) | 0.178 (0.384) | 0.558 | 0.205 (0.406) | 0.222 (0.418) | 0.042 |
| Antipurine | - | 0.001 (0.03) | 0.019 (0.136) | 0.182 | 0.026 (0.159) | 0.026 (0.159) | 0 |
| Alkalizer | - | 0.017 (0.13) | 0.047 (0.212) | 0.168 | 0.026 (0.159) | 0.034 (0.182) | 0.05 |
| Nonsteroidal anti-inflammatory drug | - | 0.026 (0.158) | 0.084 (0.279) | 0.259 | 0.111 (0.316) | 0.111 (0.316) | 0 |
| Aspirin | - | 0.001 (0.031) | 0.019 (0.136) | 0.18 | 0.009 (0.092) | 0.009 (0.092) | 0 |
| Corticosteroid | - | 0.042 (0.2) | 0.121 (0.328) | 0.294 | 0.171 (0.378) | 0.179 (0.385) | 0.022 |
| Colchicine | - | 0.0 (0.014) | 0.0 (0.0) | −0.02 | 0.0 (0.0) | 0.0 (0.0) | - |
| CKD stage | - | 2.069 (1.092) | 3.056 (1.106) | 0.898 | 3.051 (1.049) | 3.051 (1.121) | 0 |
| High-density lipoprotein cholesterol | mg/dl | 50.889 (12.33) | 49.073 (13.813) | −0.139 | 49.874 (12.093) | 49.462 (13.965) | −0.031 |
| Hemoglobin A1c | % | 6.044 (1.146) | 6.028 (1.08) | −0.015 | 5.906 (0.66) | 5.841 (0.881) | −0.083 |
| eGFR | g/dl | 86.603 (49.982) | 46.719 (28.571) | −0.98 | 47.481 (25.145) | 47.256 (29.746) | −0.008 |
| Albumin | mg/dl | 3.837 (0.676) | 3.79 (0.726) | −0.067 | 3.883 (0.757) | 3.913 (0.65) | 0.042 |
| Creatinine | mg/dl | 1.013 (1.413) | 1.896 (2.248) | 0.47 | 1.654 (1.903) | 1.77 (1.992) | 0.06 |
| Triglycerides | mg/dl | 125.427 (91.417) | 150.71 (101.225) | 0.262 | 155.222 (134.693) | 146.479 (91.511) | −0.076 |
| Serum uric acid | mg/dl | 5.087 (1.773) | 6.007 (2.322) | 0.445 | 6.153 (2.012) | 5.88 (2.498) | −0.12 |
| Age | - | 58.086 (17.471) | 65.57 (13.366) | 0.481 | 63.744 (13.141) | 64.41 (15.014) | 0.047 |
| Sex | - | 0.524 (0.499) | 0.159 (0.367) | −0.835 | 0.179 (0.385) | 0.145 (0.354) | −0.093 |
| Hemoglobin | g/dl | 12.592 (2.255) | 12.428 (2.593) | −0.068 | 12.84 (2.716) | 12.552 (2.788) | −0.105 |
| Urine pH | | 6.499 (0.601) | 6.35 (0.793) | −0.211 | 6.376 (0.576) | 6.423 (0.803) | 0.067 |
| Urine occult blood | | 0.441 (1.058) | 0.495 (1.022) | 0.052 | 0.436 (1.012) | 0.47 (0.952) | 0.035 |

IID setting

The experimental outcomes for the IID settings are shown in Fig. 2. The three graphs on the left show the metrics (gap, inconsistency, and MASMD) for experiments where allopurinol served as the treatment, whereas the three graphs on the right illustrate the results for febuxostat. In most experimental settings, using DC-QE outperformed IA, and increasing the number of collaborating users brought the results closer to CA. Under DC-QE, the performance was sometimes nearly the same as, or slightly worse than, that of a single user. However, excluding LPP, collaboration among five or more users consistently surpassed IA in all settings, with accuracy tending to improve as the number of collaborating users increased. On the other hand, when LPP was used as the dimensionality reduction function, both MASMD and inconsistency tended to worsen with each additional collaborating user (Fig. 2**c, f**). In contrast, when AE or PCA was employed, DC-QE's accuracy improved consistently as the number of collaborating users increased.

Non-IID setting

The experimental results for the non-IID setting under three data partitioning methods are presented in Figs. 3, 4, and 5: Fig. 3 for quantity-based partitioning, Fig. 4 for label-ratio partitioning, and Fig. 5 for cluster-based partitioning. IA*i* in these figures represents the *i*-th user. Figs. 3, 4, and 5 show that DC-QE outperformed IA across most metrics. However, in certain circumstances, DC-QE underperformed compared to IA. In particular, when LPP was used for dimensionality reduction, the accuracy was often worse than that of IA (Fig. 3**b**, **e, f**; Fig. 5**f**). Even with dimensionality reduction methods other than LPP, in a few cases DC-QE performed worse than IA. For example, as shown in Fig. 5**e**, when PCA was applied for dimensionality reduction, the MASMD value under DC-QE was lower than that of IA1. Furthermore, as shown in Fig. 5**f**, DC-QE underperformed IA across all experimental settings.

It was difficult to identify an optimal method for creating intermediate representations. When PCA was used, the inconsistency score was often the smallest (Fig. 3**d**; Fig. 4**c, d**; Fig. 5**c**). When bootstrap was used, MASMD was the lowest for all DC-QE settings (Fig. 3**e**, **f**; Fig. 4**e, f**; Fig. 5**e, f**). Combining bootstrap with a conventional dimensionality reduction method generally produced accuracy that fell between using only bootstrap and using only a conventional dimensionality reduction method for generating the intermediate representation.

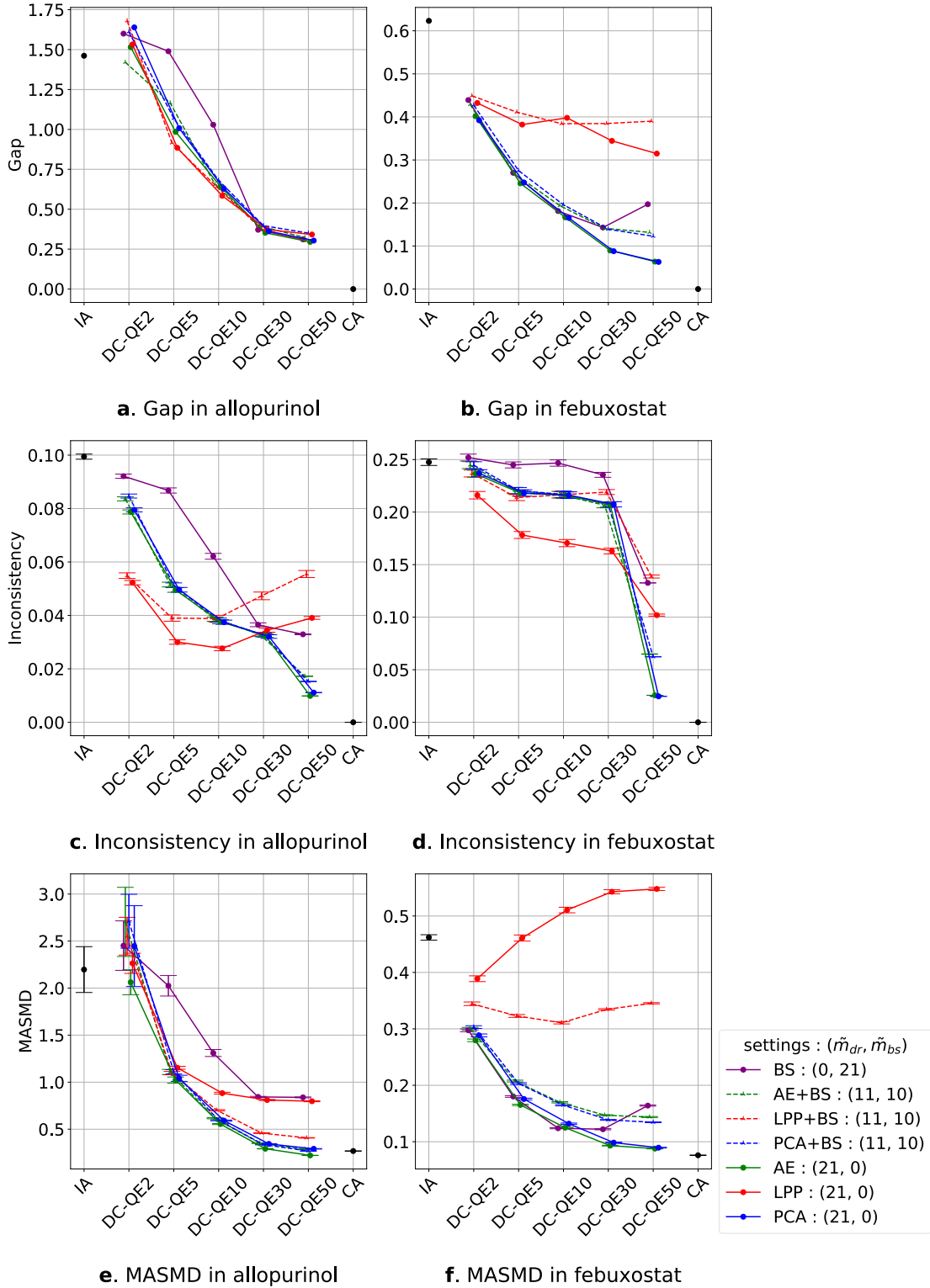

**Fig. 2.** Simulation results for the IID settings. **a, c, e:** Results where the treatment was allopurinol; **b, d, f:** results where the treatment was febuxostat. The x-axis represents collaboration settings IA, DC-QE with k collaboration (e.g., DC-QE5 refers to five users collaborating to use DC-QE), and CA. The legend shows the dimensionality reduction method and number of dimensions used to construct the dimensionality reduction function. For example, AE+BS (11,10) means autoencoder and bootstrap-based dimensionality reduction were used to construct dimensionality reduction with $(\widetilde{m}_{dr}, \widetilde{m}_{dr}) = (11,10)$. The error bar refers to the standard error (SE). Note that gap does not have SE because the gap can be calculated using whole bootstrapped results.

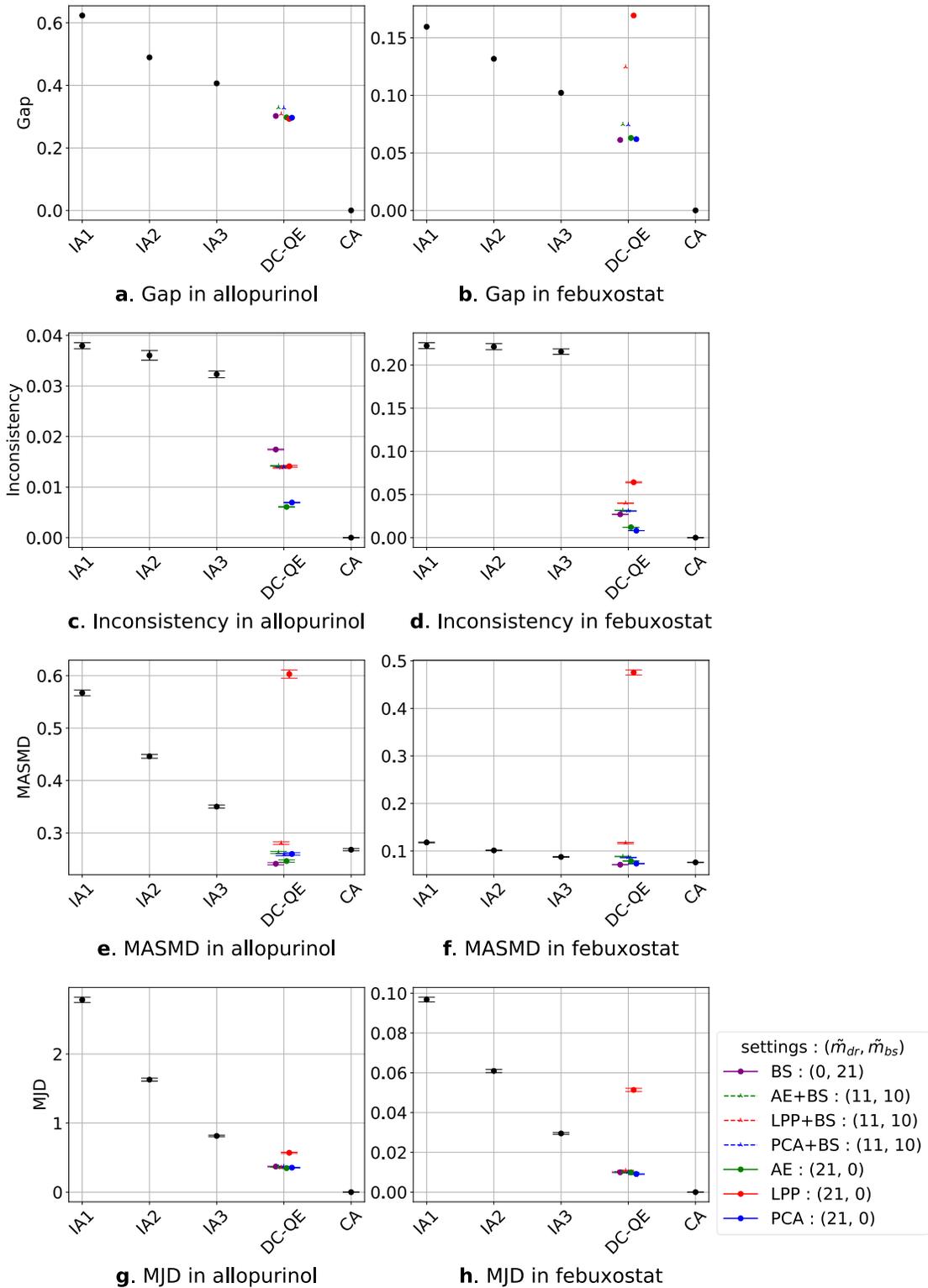

**Fig. 3.** Simulation results for the non-IID setting with quantity-based partitioning. **a, c, e, g:** Results where the treatment was allopurinol; **b, d, f, h:** results where the treatment was febuxostat. The x-axis represents the collaboration settings IA, DC-QE with k collaboration (e.g., DC-QE5 refers to five users collaborate use DC-QE), and CA. The legend shows the dimensionality reduction method and number of dimensions used to construct the dimensionality reduction function. For example, AE+BS (11,10) means autoencoder and bootstrap-based dimensionality reduction were used

to construct dimensionality reduction with $(\tilde{m}_{dr}, \tilde{m}_{dr}) = (11, 10)$. The error bar refers to the standard error (SE). Note that gap does not have SE because the gap can be calculated using whole bootstrapped results.

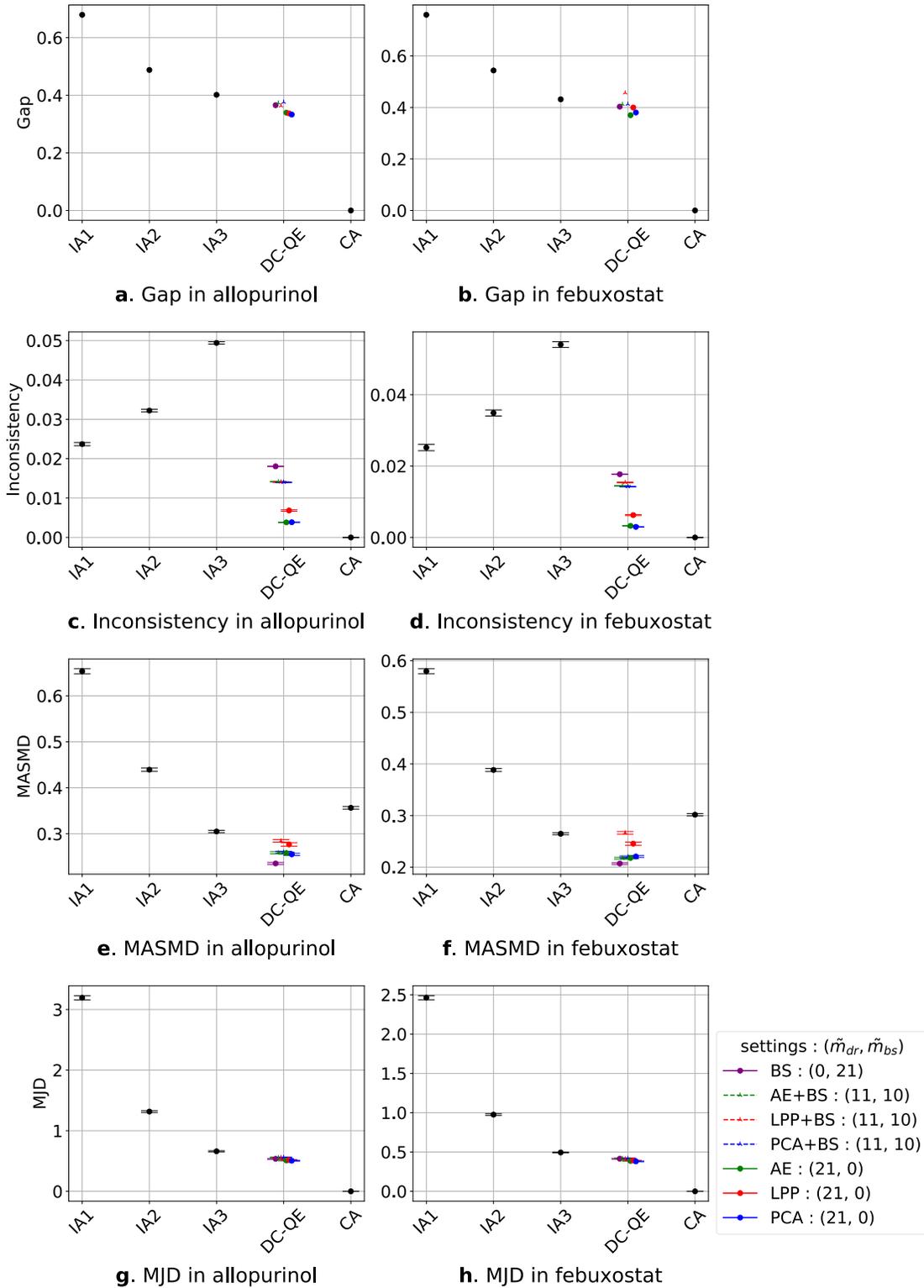

**Fig. 4.** Simulation results for the non-IID setting with label ratio-based partitioning. **a, c, e, g:** Results where the treatment was allopurinol; **b, d, f, h:** results where the treatment was febuxostat. The x-axis represents the collaboration settings, IA, DC-QE with k collaboration (e.g., DC-QE5 refers to five users collaborating to use DC-QE), and CA. The legend shows the dimensionality reduction method and number of dimensions used to construct the dimensionality reduction function. For example, AE+BS (11,10) means autoencoder and bootstrap-based dimensionality

reduction were used to construct dimensionality reduction with $(\tilde{m}_{dr}, \tilde{m}_{dr}) = (11,10)$. The error bar refers to the standard error (SE). Note that gap does not have SE because the gap can be calculated using whole bootstrapped results.

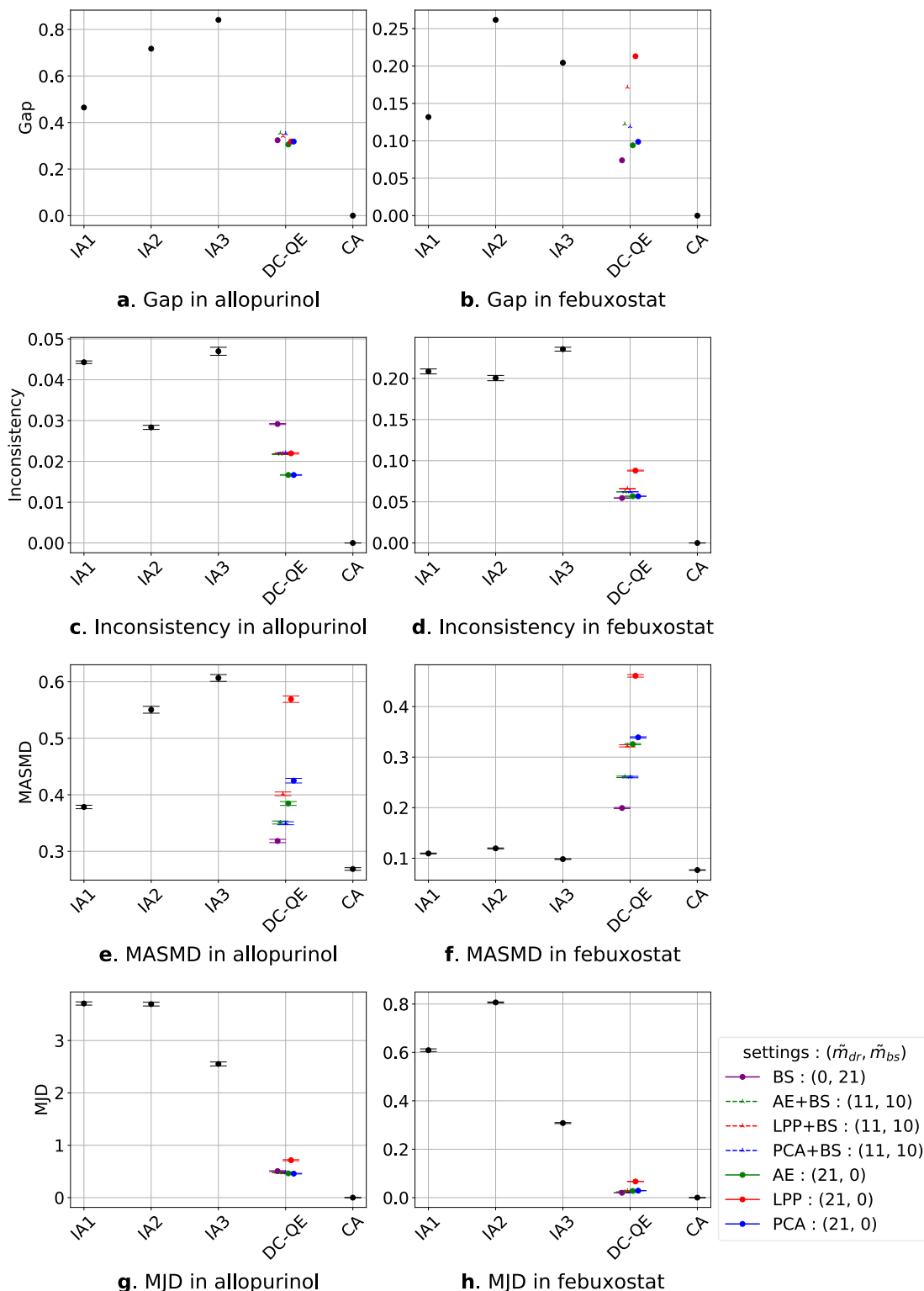

**Fig. 5.** Simulation results for the non-IID setting with cluster-based partitioning. **a, c, e, g:** Results where the treatment was allopurinol; **b, d, f, h:** results where the treatment was febuxostat. The x-axis represents the collaboration settings IA, DC-QE with k collaboration (e.g., DC-QE5 refers to five users collaborating to use DC-QE), and CA. The legend shows the dimensionality reduction method and number of dimensions used to construct the dimensionality reduction function. For example, AE+BS (11,10) means autoencoder and bootstrap-based dimensionality reduction

were used to construct dimensionality reduction with $(\widetilde{m}_{dr}, \widetilde{m}_{dr}) = (11,10)$. The error bar refers to the standard error (SE). Note that gap does not have SE because the gap can be calculated using whole bootstrapped results.

Discussion

This section discusses the results of the numerical experiments from four perspectives: the performance of DC-QE relative to IA and CA, the influence of intermediate representation methods on DC-QE performance, the potential for further performance improvements, and the limitations of the study.

Collaboration in DC-QE demonstrated superior performance over IA in most situations. The results closely approximated those of CA when all user data were included in the collaboration. Except for LPP, collaboration among five or more users consistently surpassed IA in all IID settings (Fig. 2). Moreover, improved performance with DC-QE was also observed in non-IID settings (Figs. 3, 4, and 5). These findings indicate that DC-QE exhibits robustness to medical data to some extent. Notably, performance gains were achieved even when individual users possessed small datasets. This result suggests a potential application of DC-QE in estimating treatment effects for rare diseases with limited case data.

Although the DC-QE outperformed IA in many situations, sometimes IA outperformed DC-QE. For example, when using only LPP as the dimensionality reduction function, it tended to degrade the DC-QE results (Fig. 2**f**; Fig. 3**b, e, f**; Fig. 5**b, e, f**). LPP is a dimensionality reduction method designed to preserve local structure. Because it places too much emphasis on the relationships among neighboring points, it may fail to adequately retain the global structure of the original data, resulting in information loss and potentially reducing accuracy. In addition, individual data may have different local structures, even if the data are IID partitioned, causing information loss. Except for LPP, in some settings, IA outperformed DC-QE. In cluster partitioning in non-IID settings, IA1 outperformed DC-QE in some dimensionality reduction methods (Fig. 5**e**) and IA1–IA3 outperformed DC-QE for all dimensionality reduction methods (Fig. 5**f**). This result means that the control group and treatment group were matched better in the IA population than in the DC-QE population. However, in other metrics, such as gap (Fig. 5**a, b**), inconsistency (Fig. 5**c, d**), and MJD (Fig. 5**e, f**), DC-QE outperformed IA. MJD represents the closeness of the covariate distribution to CA after matched samples, so the result of IA can be rephrased as follows: Matched accurately in an IA sample, but the matched covariate distributions in IA samples were not close to those of the CA matched population. This result may explain why IA outperformed DC-QE in terms of MASMD, even though DC-QE was superior to IA on the other metrics.

In terms of the intermediate representation method, PCA and AE exhibited better results than the other methods. However, PCA and AE are not always good options in non-IID settings. For example, in all non-IID experiments, using BS produced the best MASMD metric among the various DC-QE experimental settings (Figs. 3–5**e, f**). In addition, using BS, DC-QE had MJD scores close to CA in non-IID settings (Figs. 3–5**g, h**). These results suggest bootstrap may be suited to non-IID situations. A detailed analysis of this approach will be undertaken in future research.

The performance of DC-QE can be improved by algorithm adjustments. First, by using a low-rank adaptation method or SMOTE-based method to generate anchor data, a performance improvement can be expected. Imakura et al. [31] showed that using SMOTE improved the DC analysis, but the results for DC-QE were unclear. Second, construction of $g_k$ using another method may improve performance. In previous DC framework studies, linear functions were mainly applied to construct $g_k$. Using nonlinear functions such as the neural network-based approach may improve performance.

The primary limitation of this study was the data partitioning method. Because of ethical review constraints, only medical data from one hospital could be used, making it impossible to combine it with datasets from other hospitals. Thus, we artificially partitioned the data to create a non-IID scenario. The situation does not necessarily reflect real-world distributed situations. Thus, when actual data are combined, different results may be obtained.

## Conclusion

In this study, the application of DC-QE, a technique enabling integrated analysis and causal inference for distributed areas while preserving privacy, was demonstrated with medical datasets. Additionally, a proposed dimensionality reduction method tailored for DC-QE was evaluated using various reduction strategies.

Experimental results confirmed that DC-QE performs effectively in both IID and non-IID scenarios. Bootstrap-based dimensionality reduction proved beneficial in maintaining performance under conditions where covariate distributions varied substantially across users.

DC-QE outperformed IA even with datasets as small as 100 samples per user, highlighting its applicability in collaborative medical data analysis involving hospitals with limited data. DC-QE enables the identification of previously unknown causal relationships, such as those relevant to drug repositioning or therapeutic interventions for rare diseases, by facilitating privacy-preserving collaboration.

Future work will explore extending the DC analysis framework by combining DC-QE with survival analysis[31] to estimate hazard ratios while addressing bias. Achieving this goal would further enhance the utility of the DC framework in medical research by combining privacy preservation with robust bias elimination and hazard ratio estimation. In addition, the expansion of DC-QE for higher-dimensional data, such as covariates with high dimension or propensity scores expressed as a high-order polynomial of covariates, will be explored in future research. For example, noise-reduction methodology applied to PCA [43] may be beneficial when making intermediate representation if a user has high-dimension, low-sample-size (HDLSS) data.

## Data availability

The datasets used in this study were obtained from the University of Tsukuba Hospital and are subject to restrictions. These data were accessed under license for the current study and Japanese Act on the Protection of Personal Information, and so is not publicly available. Interested researchers may request access to the data from the corresponding author upon reasonable request and approval from the University of Tsukuba Hospital.

## Acknowledgments

The authors express their gratitude to the University of Tsukuba Hospital for providing the medical data. This study was partially funded by the Cross-Ministerial Strategic Innovation Promotion Program (SIP) on the "Integrated Health Care System" Grant (JPJ012425), the Japan Science and Technology Agency (JST) (Nos. JPMJPF2017, JPMJPR23I3 (PRESTO)), Japan Society for the Promotion of Science (JSPS), and the Grants-in-Aid for Scientific Research (Nos. JP23K28192, JP23K16249). The authors would like to thank Editage (www.editage.jp) for English-language editing.

## Author contributions

T. N., Y. K., A. T., and Y. O. designed the study. T. N. conducted data processing and analysis. T. N., Y. K., A. T. and Y. O. wrote the manuscript. Y. K., A. I., and T. S. constructed a DC analysis model tailored to the current problem set and interpreted the results mathematically. M. S. and R. T. obtained the data from the University of Tsukuba Hospital. R. K., R. T., and K. Y. assisted in validating the analysis from a medical standpoint and interpreting the results. All the authors have read and approved the final version of the manuscript.

[1] Farjat, A. E. *et al*. The importance of the design of observational studies in comparative effectiveness research: Lessons from the GARFIELD-AF and ORBIT-AF registries. *Am. Heart J*. **243**, 110–121 (2022).

[2] Bärnighausen, T. *et al*. Quasi-experimental study designs series-paper 4: Uses and value. Quasi-Experimental Study Designs Series—paper 4. *J. Clin. Epidemiol*. **89**, 21–29 (2017).

[3] Ligthelm, R. J. *et al*. Importance of observational studies in clinical practice. *Clin. Ther*. **29**, 1284–1292 (2007).

[4] Wang, Y.-R. & Tsai, Y.-C. The protection of data sharing for privacy in financial vision. *Appl. Sci*. **12**, 7408 (2022).

[5] Simpson, E. *et al*. Understanding the barriers and facilitators to sharing patient-generated health data using digital technology for people living with long-term health conditions: A narrative review. *Front. Public Health* **9**, 641424 (2021).

[6] El Emam, K., Rodgers, S. & Malin, B. Anonymising and sharing individual patient data. *BMJ* **350**, h1139–h1139 (2015).

[7] Hulsen, T. Sharing is Caring-Data sharing initiatives in healthcare. *Int. J. Environ. Res. Public Health* **17**, 3046 (2020).

[8] McMahan, B., Moore, E., Ramage, D., Hampson, S. & Arcas, B. A. y. Communication-efficient learning of deep networks from decentralized data in *Proceedings of the 20th International Conference on Artificial Intelligence and Statistics 1273–1282 (PMLR, 2017)*.

[9] Rieke, N. *et al*. The future of digital health with federated learning. *npj Digit. Med*. **3**, 119 (2020).

[10] Ashish Rauniyar *et al*. Federated learning for medical applications: A taxonomy, current trends, challenges, and future research directions, *Ar5iv*. https://ar5iv.labs.arxiv.org/html/2208.03392 (2022).

[11] Crowson, M. G. *et al*. A systematic review of federated learning applications for biomedical data. *PLOS Digit. Health* **1**, e0000033 (2022).

[12] Prayitno, *et al*. A systematic review of federated learning in the healthcare area: From the perspective of data properties and applications. *Appl. Sci*. **11**, 11191 (2021).

[13] Imakura, A. & Sakurai, T. Data collaboration analysis framework using centralization of individual intermediate representations for distributed data sets. *ASCE ASME J. Risk Uncertainty Eng. Syst. A* **6** (2020).

[14] Uchitachimoto, G. *et al.* Data collaboration analysis in predicting diabetes from a small amount of health checkup data. *Sci. Rep.* **13**, 11820 (2023).

[15] Kawamata, Y., Motai, R., Okada, Y., Imakura, A. & Sakurai, T. Collaborative causal inference on distributed data. *Expert Syst. Appl.* **244**, 123024 (2024).

[16] Rosenbaum, P. R. & Rubin, D. B. The central role of the propensity score in observational studies for causal effects. *Biometrika* **70**, 41–55 (1983).

[17] Zeng, R. *et al.* Associations of proton pump inhibitors with susceptibility to influenza, pneumonia, and COVID-19: Evidence from a large population-based cohort study. *eLife* **13**, RP94973 (2024).

[18] Hsu, Y.-C. *et al.* Comparing lenvatinib/pembrolizumab with Atezolizumab/Bevacizumab in unresectable hepatocellular carcinoma: A real-world experience with propensity score matching analysis. *Cancers* **16**, 3458 (2024).

[19] Yang, Q. *et al.* Ischemic cardio-cerebrovascular disease and all-cause mortality in Chinese elderly patients: A propensity-score matching study. *Eur. J. Med. Res.* **29**, 330 (2024).

[20] Chuang, M.-H. *et al.* New-onset obstructive airway disease following COVID-19: A multicenter retrospective cohort study. *BMC Med.* **22**, 360 (2024).

[21] Vo, T. V., Lee, Y., Hoang, T. N. & Leong, T. Y. Bayesian federated estimation of causal effects from observational data. In *Uncertainty in Artificial Intelligence*, 2024–2034 (PMLR, 2022).

[22] Xiong, R., Koenecke, A., Powell, M., Shen, Z., Vogelstein, J. T. & Athey, S. Federated causal inference in heterogeneous observational data. *Stat. Med.* **42**, 4418–4439 (2023).

[23] Han, L., Li, Y., Niknam, B. & Zubizarreta, J. R. Privacy-preserving, communication-efficient, and target-flexible hospital quality measurement. *Ann. Appl. Stat.* **18**, 1337–1359 (2024).

[24] Iseki, K., Horio, M., Imai, E., Matsuo, S. & Yamagata, K. Geographic difference in the prevalence of chronic kidney disease among Japanese screened subjects: Ibaraki versus Okinawa. *Clin. Exp. Nephrol.* **13**, 44–49 (2009).

[25] Chou, H.-W. *et al.* Comparative effectiveness of allopurinol, febuxostat and benzbromarone on renal function in chronic kidney disease patients with hyperuricemia: A 13-year inception cohort study. *Nephrol. Dial. Transplant.* **33**, 1620–1627 (2018).

[26] Becker, M. A. *et al.* Febuxostat compared with allopurinol in patients with hyperuricemia and gout. *N. Engl. J. Med.* **353**, 2450–2461 (2005).

[27] Bruce S. P. Febuxostat: A Selective Xanthine Oxidase Inhibitor for the Treatment of Hyperuricemia and Gout - Susan P Bruce, 2006. *The Annals of pharmacotherapy* **40**, 2187–2194 (2006).

[28] Hosoya, T, Sasaki, T. & Ohashi, T. Clinical efficacy and safety of topiroxostat in Japanese hyperuricemic patients with or without gout: a randomized, double-blinded, controlled phase 2b study. *Clin. Rheumatol.* 36, 649–656 (2017).

[29] Matsuo, S. *et al.* Revised equations for estimated GFR from serum creatinine in Japan. *Am. J. Kidney Dis.* **53**, 982–992 (2009).

[30] Imakura, A., Inaba, H., Okada, Y. & Sakurai, T. Interpretable collaborative data analysis on distributed data. *Expert Syst. Appl.* **177**, 114891 (2021).

[31] Imakura, A., Kihira, M., Okada, Y. & Sakurai, T. Another use of SMOTE for interpretable data collaboration analysis. *Expert Syst. Appl.* **228**, 120385 (2023).

[32] Wold, S., Esbensen, K. & Geladi, P. Principal component analysis. *Chemom. Intell. Lab. Syst.* **2**, 37–52 (1987).

[33] Fisher, R. A. The use of multiple measurements in taxonomic problems. *Ann. Eugen.* **7**, 179–188 (1936).

[34] He, X. & Niyogi, P. Locality preserving projections in *Adv. Neural Inf. Process. Syst.* (MIT Press, 2003) **16**.

[35] Imakura, A., Bogdanova, A., Yamazoe, T., Omote, K. & Sakurai, T. Accuracy and privacy evaluations of collaborative data analysis. In *Proc. Second AAAI Workshop on Privacy-Preserving Artificial Intelligence (PPAI-21)* (2021).

[36] Nguyen, H., Zhuang, D., Wu, P.-Y. & Chang, M. AutoGAN-based dimension reduction for privacy preservation. *Neurocomputing* **384**, 94–103 (2020).

[37] Imakura, A., Tsunoda, R., Kagawa, R., Yamagata, K. & Sakurai, T. DC-COX: Data collaboration Cox proportional hazards model for privacy-preserving survival analysis on multiple parties. *J. Biomed. Inform.* **137**, 104264 (2023)

[38] Andrillon, A., Pirracchio, R. & Chevret, S. Performance of propensity score matching to estimate causal effects in small samples. *Stat. Methods Med. Res*. **29**, 644–658 (2020).

[39] Austin, P. C. A critical appraisal of propensity-score matching in the medical literature between 1996 and 2003. *Stat. Med.* **27**, 2037–2049 (2008).

[40] Kane, L. et al. Propensity score matching: A statistical method. *Clin. Spine Surg.* **33**, 120-122 (2020).

[41] Rumelhart, D. E., Hinton, G. E. & Williams, R. J. Learning internal representations by error propagation. In *Parallel Distributed Processing: Explorations in the Microstructure of Cognition, Vol. 1: Foundations*, 318–362 (MIT Press, 1986).

[42]. Austin, P. C. & Stuart, E. A. Moving towards best practice when using inverse probability of treatment weighting (IPTW) using the propensity score to estimate causal treatment effects in observational studies. *Stat. Med.* **34**, 3661–3679 (2015).

[43]. Jeffreys, H. An invariant form for the prior probability in estimation problems. *Proc. R. Soc. Lond. A Math. Phys. Sci.* **186**, 453–461 (1946).

[43]. Yata, K. & Aoshima, M. Effective PCA for high-dimension, low-sample-size data with noise reduction via geometric representations. *J. Multivar. Anal.* **105**, 193–215 (2012).

**Figure legends**

**Fig. 1.** Preprocessing flowchart, where *n* indicates the number of samples.

**Fig. 2.** Simulation results for the IID settings. **a, c, e:** Result*s* where the treatment was allopurinol; **b, d, f:** result*s* where the treatment was febuxostat. The x-axis represents collaboration settings IA, DC-QE with k collaboration (e.g., DC-QE5 refers to five users collaborating to use DC-QE), and CA. The legend shows the dimensionality reduction method and number of dimensions used to construct the dimensionality reduction function. For example, AE+BS (11,10) means autoencoder and bootstrap-based dimensionality reduction were used to construct dimensionality reduction with $(\widetilde{m}_{dr}, \widetilde{m}_{dr}) = (11,10)$. The error bar refers to the standard error (SE). Note that gap does not have SE because the gap can be calculated using whole bootstrapped results.

**Fig. 3.** Simulation results for the non-IID setting with quantity-based partitioning. **a, c, e, g:** Results where the treatment was allopurinol; **b, d, f, h:** results where the treatment was febuxostat. The x-axis represents the collaboration settings IA, DC-QE with k collaboration (e.g., DC-QE5 refers to five users collaborate use DC-QE), and CA. The legend shows the dimensionality reduction method and number of dimensions used to construct the dimensionality reduction function. For example, AE+BS (11,10) means autoencoder and bootstrap-based dimensionality reduction were used to construct dimensionality reduction with $(\widetilde{m}_{dr}, \widetilde{m}_{dr}) = (11,10)$. The error bar refers to the standard error (SE). Note that gap does not have SE because the gap can be calculated using whole bootstrapped results.

**Fig. 4.** Simulation results for the non-IID setting with label ratio-based partitioning. **a, c, e, g:** Results where the treatment was allopurinol; **b, d, f, h:** results where the treatment was febuxostat. The x-axis represents the collaboration settings IA, DC-QE with k collaboration (e.g., DC-QE5 refers to five users collaborating to use DC-QE), and CA. The legend shows the dimensionality reduction method and number of dimensions used to construct the dimensionality reduction function. For example, AE+BS (11,10) means autoencoder and bootstrap-based dimensionality reduction were used to construct dimensionality reduction with $(\widetilde{m}_{dr}, \widetilde{m}_{dr}) = (11,10)$. The error bar refers to the standard error (SE). Note that gap does not have SE because the gap can be calculated using whole bootstrapped results.

**Fig. 5.** Simulation results for the non-IID setting with cluster-based partitioning. **a, c, e, g:** Results where the treatment was allopurinol; **b, d, f, h:** results where the treatment was febuxostat. The x-axis represents the collaboration settings IA, DC-QE with k collaboration (e.g., DC-QE5 refers to five users collaborating to use DC-QE), and CA. The legend shows the dimensionality reduction method and number of dimensions used to construct the dimensionality reduction function. For example, AE+BS (11,10) means autoencoder and bootstrap-based dimensionality reduction were used to construct dimensionality reduction with $(\widetilde{m}_{dr}, \widetilde{m}_{dr}) = (11,10)$. The error bar refers to the standard error (SE). Note that gap does not have SE because the gap can be calculated using whole bootstrapped results.